\documentclass[final]{aipproc}

\layoutstyle{8x11single}

\SetInternalRegister\hbadness{8000} 

\newcommand\doingARLO[2][]{%
  \ifx\mmref\undefined #1\else #2\fi
}

\def\be{\begin{equation}}
\def\ee{\end{equation}}
\def\bea {\begin{eqnarray}}
\def\eea {\end{eqnarray}}

\begin{document}

\title{Multiplicity Distributions and Charged-Neutral Fluctuations}

\author{Tapan K. Nayak}{
 address={(for the WA98 Collaboration)}
}

\maketitle

\noindent
M.M.~Aggarwal$^{4}$,   
A.L.S.~Angelis$^{7}$,  
V.~Antonenko$^{13}$,   
V.~Arefiev$^{6}$,   
V.~Astakhov$^{6}$,  
V.~Avdeitchikov$^{6}$,   
T.C.~Awes$^{16}$,  
P.V.K.S.~Baba$^{10}$,
S.K.~Badyal$^{10}$,   
S.~Bathe$^{14}$
% ...
\\

\noindent
% ...
$^4$University of Panjab, Chandigarh 160014, India   
% ...
$^6$Joint Institute for Nuclear Research, RU-141980 Dubna, Russia  
$^7$University of Geneva, CH-1211 Geneva 4,Switzerland   
% ...
$^{10}$University of Jammu, Jammu 180001, India   
% ...
$^{13}$RRC ``Kurchatov Institute'', RU-123182 Moscow, Russia  
$^{14}$University of M{\"u}nster, D-48149 M{\"u}nster, Germany 
% ...
$^{16}$Oak Ridge National Laboratory, Oak Ridge, Tennessee 37831-6372, USA 
% ...

\bigskip
\bigskip

\begin{minipage}[h]{6.1in}
Abstract: We discuss the importance of event-by-event fluctuations in relativistic
heavy-ion collisions. We present results of multiplicity fluctuations
of photons and charged particles produced in 158$\cdot A$ GeV Pb+Pb collisions
at CERN-SPS. Multiplicity fluctuations agree with participant model calculations
indicating the absence of critical fluctuation. Localized fluctuations in
charged particles to photons have been analyzed and compared to various mixed
events. The results show localized fluctuations in both charged particles and
photons. However, no correlated DCC-like fluctuation is observed.
\end{minipage}

\bigskip

The primary goal of ultra-relativistic heavy-ion collisions is to investigate
the properties of matter at high energy density and high temperature. These studies
provide information on the dynamics of multiparticle production mechanism and
the phase transition from hadronic matter to Quark-Gluon plasma (QGP). One of the major
advantages of heavy-ion collisions at high energies is that most of the observables
can be studied in an event-by-event (E-by-E) basis because of the production of large number 
of particles in every event. This is particularly important as one expects large
fluctuations in several observables associated with the phase transition.

The subject of E-by-E fluctuations has recently gained considerable interest primarily
motivated by the near perfect Gaussian distributions of mean transverse momentum and
particle ratios \cite{na49} obtained by the NA49 experiment at CERN-SPS. 
As we will see later in this manuscript the multiplicity
distributions for narrow centrality bins are also near perfect Gaussians. The variance of
the width of these distributions contain information about the reaction mechanism as well
as the nuclear geometry. The task here is to distinguish between
statistical fluctuations and those which have dynamical origin. Several methods have
been put forward which suggest ways to infer about the presence of dynamical fluctuations
which have new physics origin. These are done by
evaluating the effect of nuclear geometry,
influence of hadronic resonances and rescattering, and contributions from Bose-Einstein 
correlations \cite{mrow,step,asakawa,jeon,heisel}. Primary interest would be
to identify critical fluctuations associated with the QGP phase transition.

Another interesting phenomenon is the formation of disoriented chiral condensates
(DCC) \cite{dcc}, which is a consequence of chiral phase transition. It gives
rise to large E-by-E fluctuations in the number of charged
particles and photons in localized phase space. The distributions of
correlation functions using charged particles and photons have been shown to be
near perfect Gaussians in the absence of DCC. The presence of DCC tends to make the
distributions wider as has been shown in ref. \cite{nandi}.

In order to infer about the presence of non-statistical fluctuations, one has to
compare the distributions obtained from experimental data to known models which
incorporate all the known phenomena. An alternate or may be complimentary procedure
to probe these fluctuations in a model independent manner would be to compare
data distributions to those of mixed events
generated from the data. Once properly understood, mixed events provide the most
unambiguous means to probe any new physics. In this manuscript we present results
from Pb+Pb collisions at $158\cdot A$GeV taken at CERN-SPS by the WA98 experimental
setup. The results will be compared to a participant model and also to several
different types of mixed events.

In the WA98 experiment, the emphasis has been on high precision, simultaneous
detection of both photons and hadrons. Photon multiplicities ($N_{\gamma-{\rm like}}$) were
measured by the preshower photon multiplicity detector (PMD) placed at 21.4 meters from the
target and covering an $\eta$ range of $2.9-4.2$. Charged particle hits ($N_{\rm ch}$) were 
counted using a circular silicon pad detector (SPMD) located 32.8~cm from the target with a
coverage of $2.35<\eta<3.75$. The centrality of the interaction is determined by the total
transverse energy ($E_{\rm T}$) measured in the mid rapidity calorimeter. The centralities 
are expressed as fractions of the measured cross section from the minimum bias $E_{\rm T}$
distributions.

Figure~1 shows the minimum bias distributions of $N_{\gamma-{\rm like}}$ clusters and 
$N_{\rm ch}$. Superimposed to these are the distributions corresponding to the centrality
cuts of top 1\%, 2\% and 5\% of the minimum bias $E_{\rm T}$ distributions. These distributions
turn out to be near perfect Gaussians with $\chi^2/ndf$ close to unity. The solid lines show 
the Gaussian fits to these curves. It has been observed
that the distributions deviate from Gaussians if the centrality bin is made broader than $0-5$\%.
In order to study the centrality dependence of fluctuations we have chosen 
2\% bins in centrality, $viz.$, 0-2\%, 2-4\%,...,58-60\%,.. etc. The resulting
multiplicity distributions are near perfect Gaussians over large centrality range, from
peripheral to central collisions.
\begin{figure}[htpb]
\includegraphics[scale=0.4]{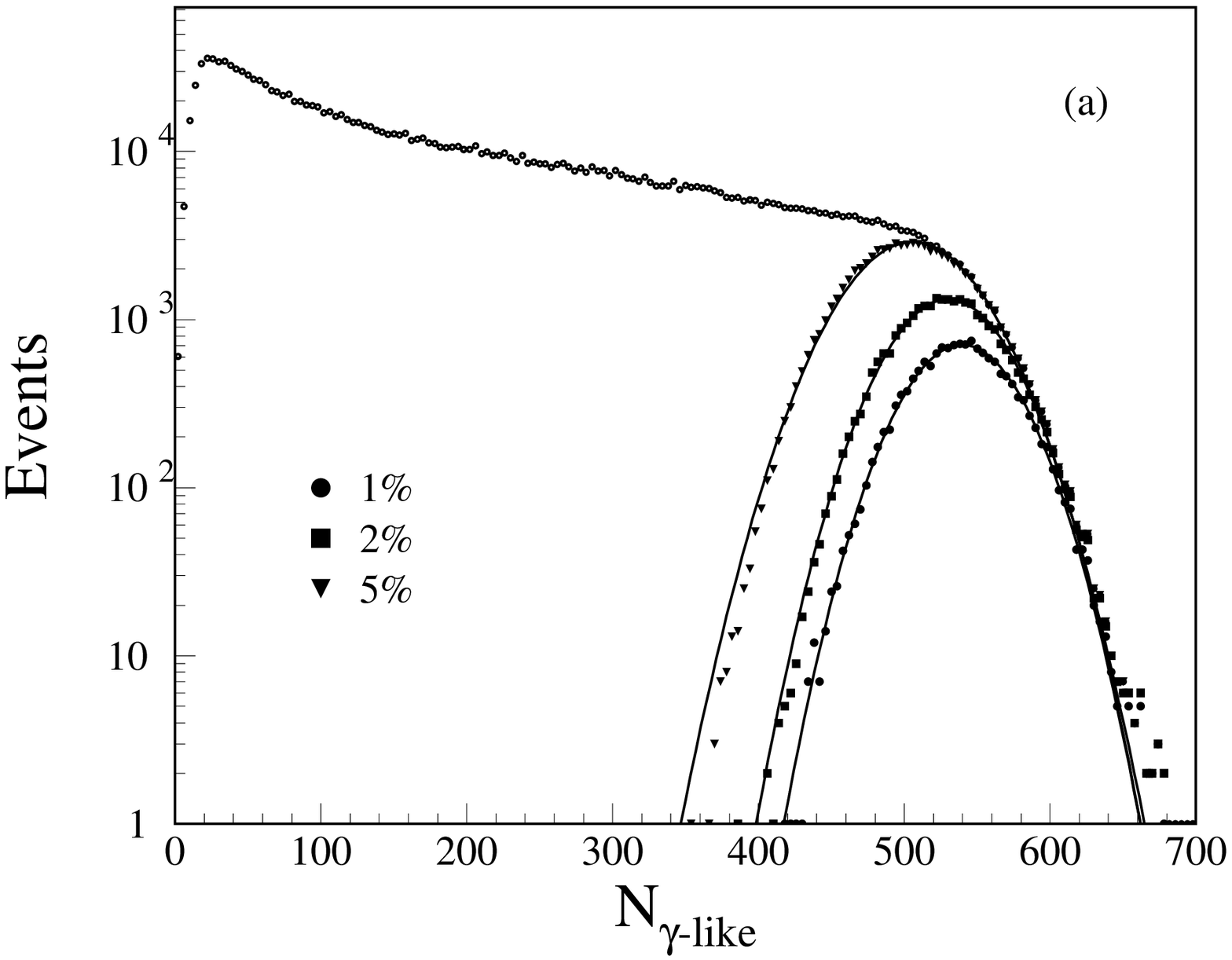}
\includegraphics[scale=0.4]{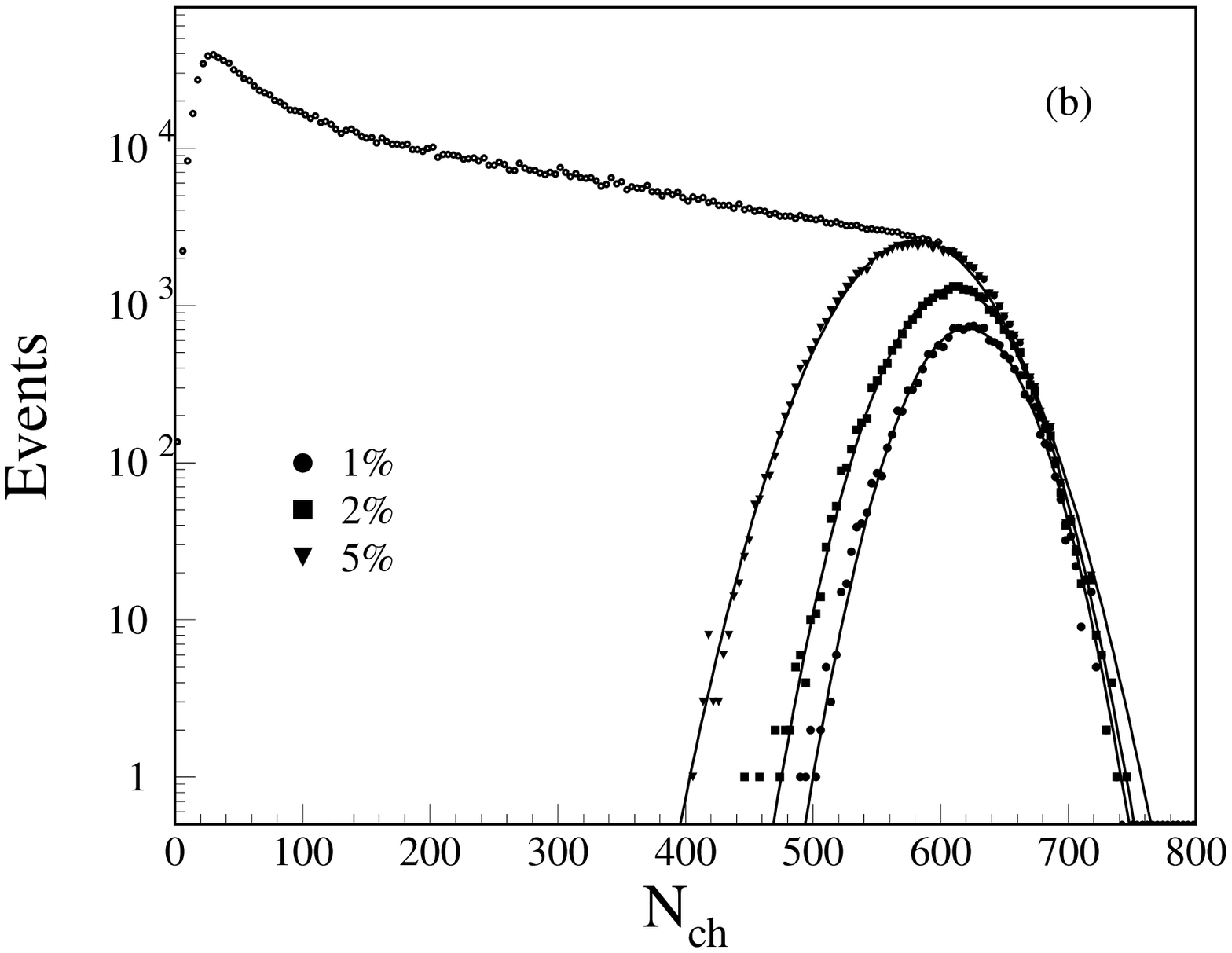}
\caption{ Minimum bias distributions of 
(a) $\gamma$-like clusters and
(b) charged particle multiplicity produced 
    in Pb induced reactions at 158$\cdot A$ GeV on Pb.
Superimposed are the multiplicity distributions for the top $1\%$, $2\%$, and $5\%$
most central events. These distributions turn out to be near perfect Gaussians as
seen by the solid fitted lines.
}
\label{nch_ngam}
\end{figure}

If the distribution of a quantity, $X$, is Gaussian, then the amount of fluctuation
may be defined as:
\begin{equation}
 \omega_X = \frac{\sigma_X^2}{\langle X \rangle},
\end{equation}
where $\sigma_x$ is the variance of the distribution. and $\langle X \rangle$ denotes the
mean value of $X$. In order to compare the fluctuation for peripheral to central collisions,
we have calculated the relative fluctuation defined as 
$\omega_X$ per number of participant nuclei, $N_{\rm part}$ (obtained
from VENUS event generator). 

Relative multiplicity fluctuations of photons ($\omega_{\gamma}/N_{\rm part}$) and
charged particles
($\omega_{\rm ch}/N_{\rm part}$), as functions of $N_{\rm part}$ are shown in left and right
panels of Figure~2, respectively. Since the measured data for photons
contain contaminations from charged particles, these have been corrected to obtain
distributions for photons only by using the efficiency and purity of the measured
photon-like clusters \cite{WA98-9}.
The resulting data values have been compared with simple
calculations based on participant model \cite{heisel,WA98-15} and also those obtained from
VENUS event generator. A reasonable agreement have been seen with the participant model
as well as VENUS indicating the absence of any critical fluctuations. 
\begin{figure}[htpb]
\includegraphics[scale=0.4]{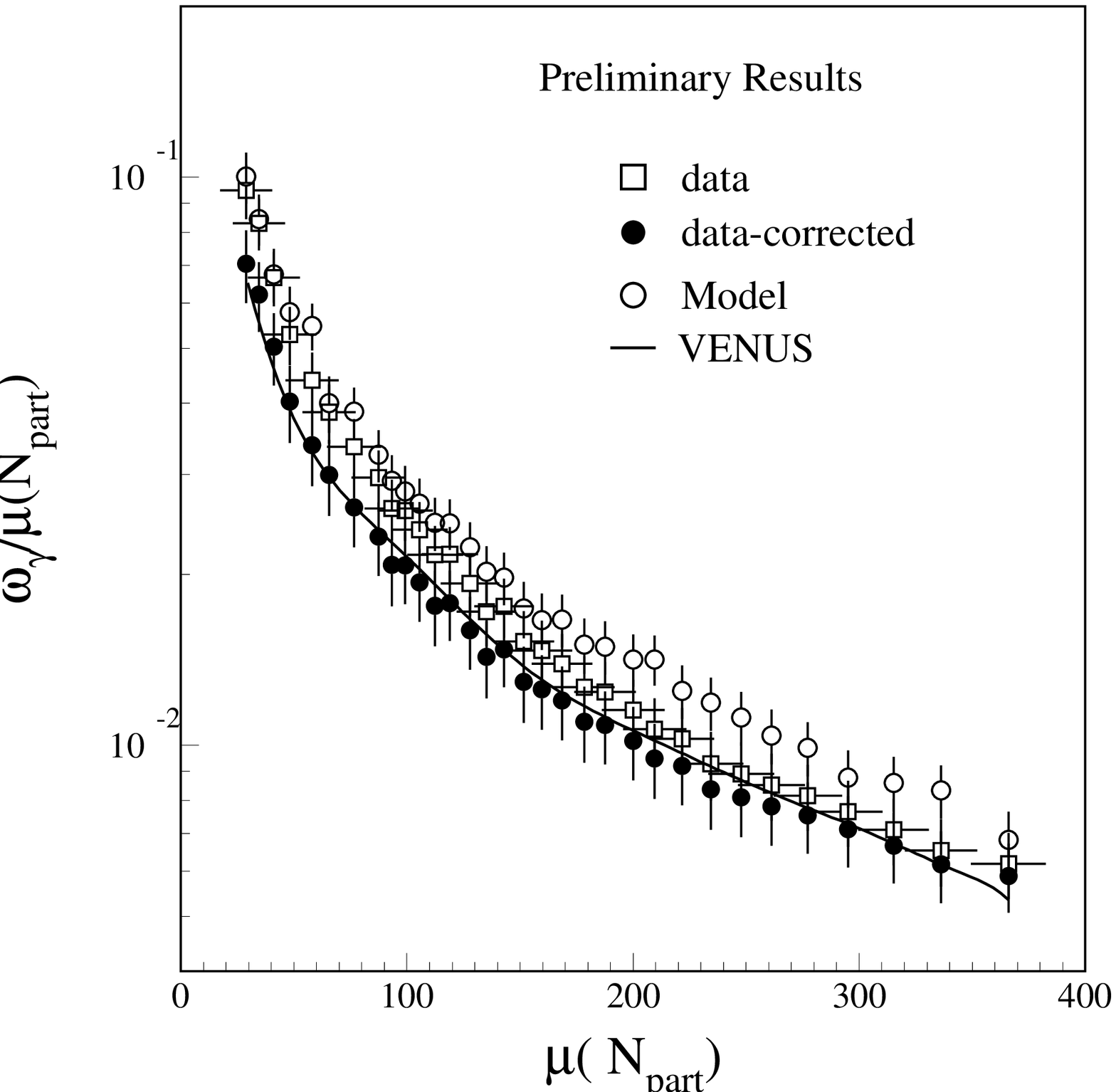}
\includegraphics[scale=0.4]{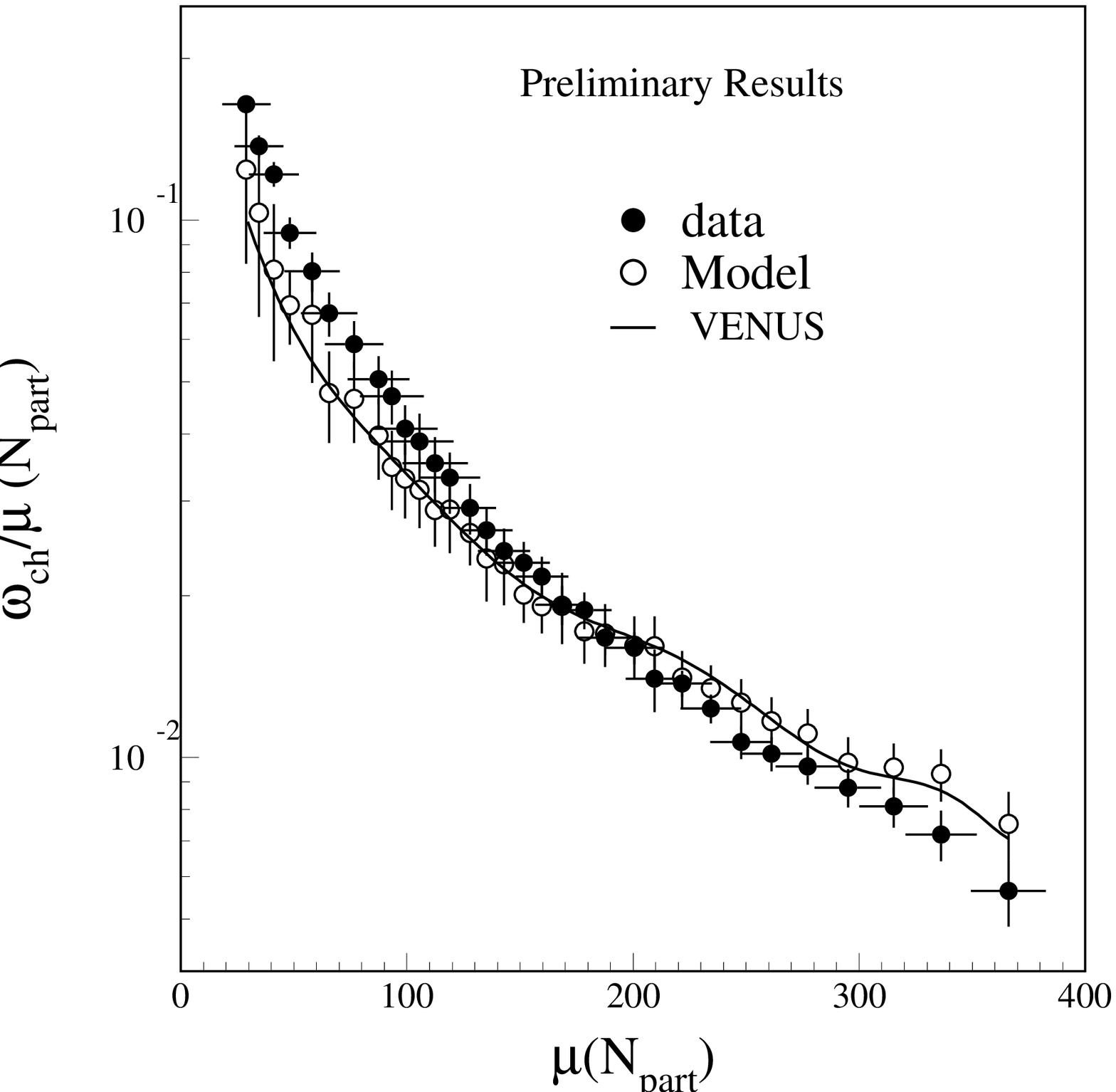}

\bigskip
\caption {\label{ch_fluc_model}
The relative fluctuations of (a) photons and (b) the charged particles
compared to calculations from a participant model and VENUS.
The photon data have been corrected from $N_{\gamma-{\rm like}}$ to
$N_{\gamma}$ using the efficiency and purity of the measured photon
sample. Reasonable agreement has been seen between data and models.
}
\end{figure}

Theoretical predictions suggest that the isospin fluctuations caused by DCC would
produce clusters of coherent pions in phase space forming domains localized in phase
space. We present results based on E-by-E fluctuation in the relative number of
charged particles and photons. Details may be obtained from Ref. \cite{WA98-12}.
Two different methods are used for DCC search. The first method deals with the
correlation of $N_{\gamma-{\rm like}}$ and $N_{\rm ch}$ and the second method
is based on the discrete wavelet techniques (DWT).

The correlation of $N_{\gamma-{\rm like}}$ and $N_{\rm ch}$ has been studied in smaller
$\phi$-segments by dividing the $\phi$-space into 2,4,8 and 16 bins. A common correlation
axis ($Z$) has been obtained by fitting the distribution of the correlations with a second
order polynomial. The distribution of the closest distance of the data points to the 
correlation axis represents the relative fluctuation of  $N_{\gamma-{\rm like}}$ 
and $N_{\rm ch}$ for any given $\phi$ bin. The final distributions ($S_{\rm Z}$) for
4 and 8 bins are shown in the left-top panel of Figure~3. Superimposed are the distributions
from GEANT simulated events and a particular set of mixed events, called M1.

We have also used the DWT method where the analysis is performed by dividing the azimuthal
space of PMD and SPMD into $2^j$ bins where $j$ is called the scale of the division. 
The input to the DWT analysis is a spectrum of the function, 
$N_{\gamma_{\rm like}}/(N_{\gamma_{\rm like}} + N_{\rm ch})$, at the highest resolution
scale, $j_{\rm max}(=5)$. The output consists of a set of wavelet or the so called father function
coefficients (FFC) at each scale, from $j=1$ to $j=j_{\rm max}-1$. The FFC distributions at
4 and 8 bins in $\phi$ are shown in the left-bottom panel of Figure~3.
The $S_{\rm Z}$ and FFC distributions are Gaussian in nature in the absence of DCC.
The presence of events with DCC domains of a particular bin size in $\phi$
would result in a broader distribution of $S_{\rm Z}$ and FFC distributions compared to
normal events \cite{nandi,WA98-12}.

The presence of DCC-like fluctuations in the experimental data can be inferred by comparing
the data distributions to different sets of mixed events. Properly constructed mixed events
preserve all the detector effects while removing correlations. Four sets of mixed events are
generated by maintaining the global (bin 1) $N_{\gamma_{\rm like}}$-$N_{\rm ch}$
correlations as in the real data. Table~1 summarizes the construction of mixed events
and the type of fluctuation these can probe. These behavior has been understood
by simulating the effect of DCC and analyzing the simulated data in a manner identical to
the real data.
\begin{table}
\begin{tabular}{|lccc|}\hline
             & PMD      & SPMD       & Type of fluctuation to probe: \\ \hline
M1           & Mix hits & Mix hits   & Correlated + Individual \\ 
M2           & ~~~~~~~~~~~~Unaltered hits~~~~~~~~~~~~ &  ~~~~~~~~~~~~Unaltered hits~~~~~~~~~~~~   & Correlated \\ 
M3-$\gamma$  &  Unaltered hits &  Mix hits  & $N_{\gamma}$ only \\ 
M3-ch        &  Mix hits &    Unaltered hits  & $N_{\rm ch}$ only \\ \hline
\end{tabular}
\caption{Four different types of mixed events are constructed by combinations
         of PMD and SPMD hits. The last column provides the usefulness of each of the
         mixed events to probe different type of fluctuation.}
\end{table}

\begin{figure}[h]
\includegraphics[scale=0.5]{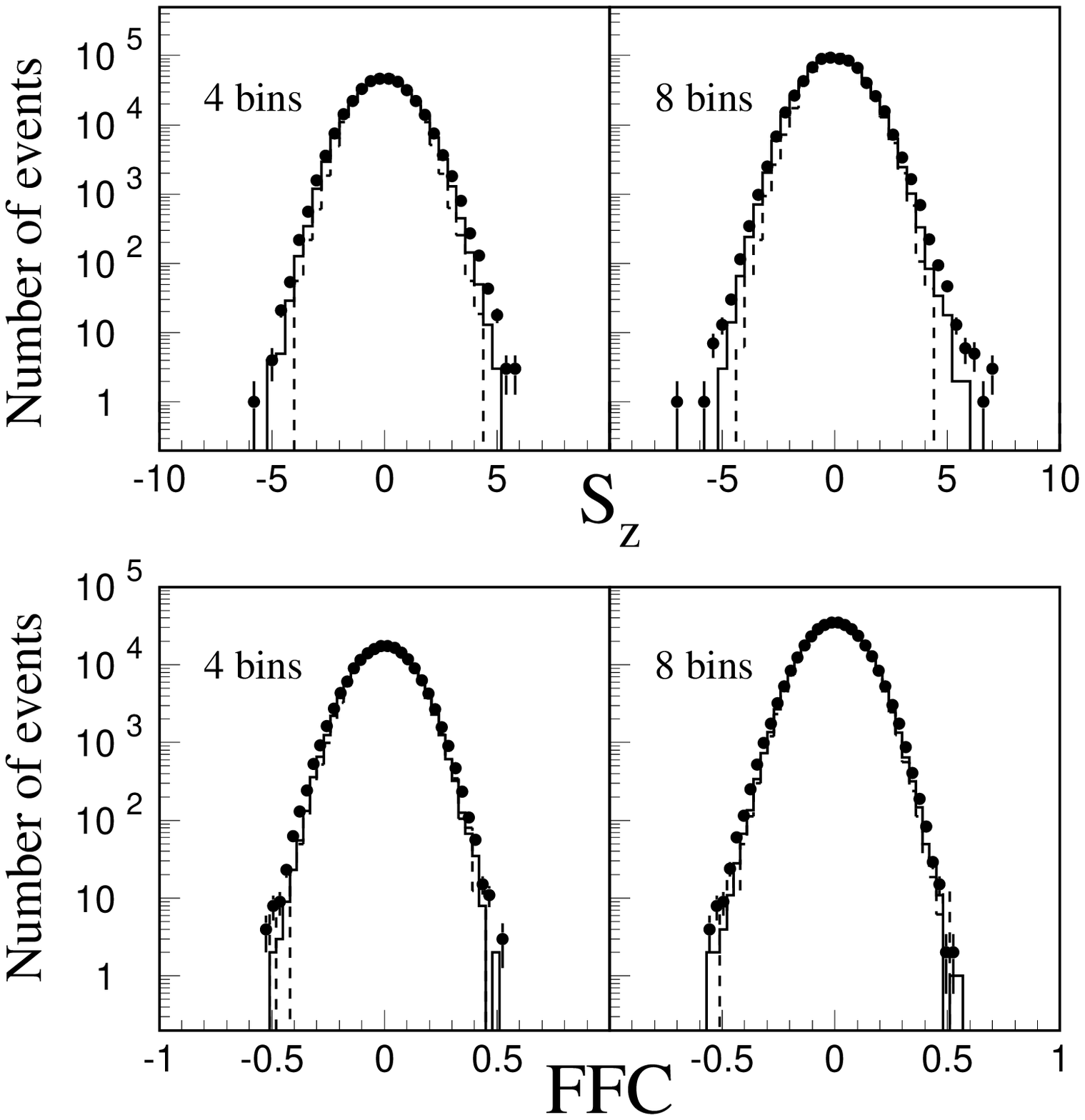}
\includegraphics[scale=0.4]{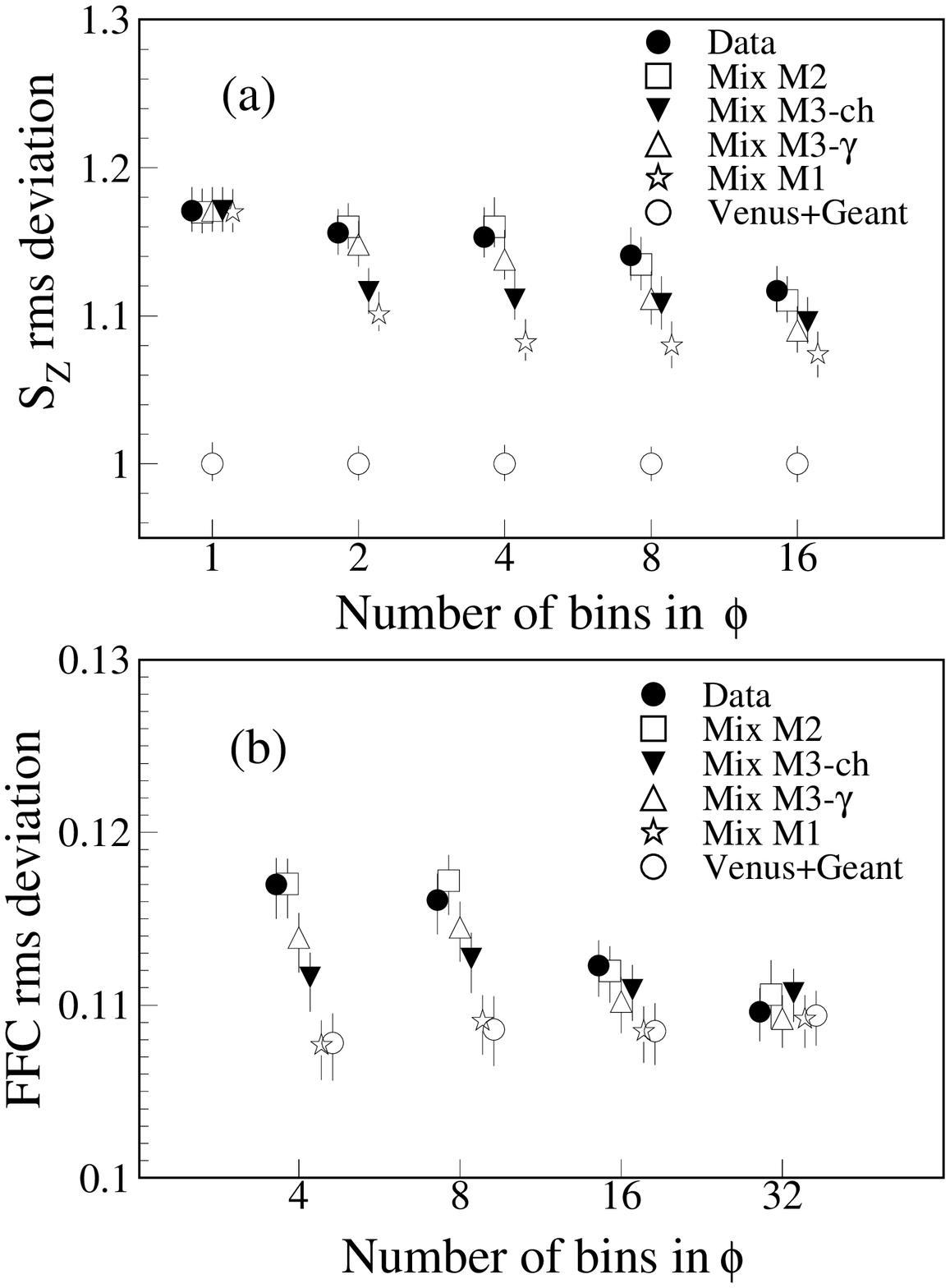}
\caption {\label{sz-ffc}
The $S_Z$ and FFC distributions for 4 and 8 divisions in $\phi$ are shown
in the left panel.
The experimental data, M1 and V+G events are
shown by solid circles, solid histograms and dashed histograms,
respectively.
The statistics for data and mixed events are the same, whereas the
distribution for the V+G events is normalized to the number of data
events. The right panel shows the root mean square (rms) deviations
of the $S_{\rm Z}$ and FFC distributions for various divisions in
the azimuthal angle.
}
\end{figure}

The rms deviations of the $S_{\rm Z}$ and FFC distributions as a function of the number of
bins in azimuth are shown for experimental data, four types of mixed events and Geant
simulated data in the right panel of Figure~3. The statistical errors are small and the
systematic errors are shown in the figure. \\

First we compare the rms deviations obtained from the data to those of M1 mixed events.
The data points are much larger compared to those of M1 for both
$S_{\rm Z}$ and FFC distributions. For 2,4 and 8 bins the values of the $S_{\rm Z}$
rms deviations of data are 2.5$\sigma$, 3.0$\sigma$ and 2.4$\sigma$ larger than those of
the M1 events. Similarly for 4 and 8 bins the values of the FFC
rms deviations of data are 3.7$\sigma$ and 2.8$\sigma$ larger than those of
the M1 events. These indicate the presence of localized non-statistical fluctuations.
The observed effect may arise because of (a) E-by-E correlated $N_{\gamma-{\rm like}}$-$N_{\rm ch}$ 
fluctuations, (b) fluctuations in $N_{\gamma-{\rm like}}$, and (c) fluctuations in $N_{\rm ch}$.
The source of the fluctuation can be obtained by comparing the data with M2 and M3 types of
events.

Next we compare data rms deviations to those of M2.
It is observed that in all cases the results from M2 match completely with data.
This implies the absence of any E-by-E correlated (DCC-like) fluctuations and allows
us only to put an upper limit on DCC production \cite{WA98-12}.

Presence of individual fluctuations are to be inferred by comparing data with M3 type of
events. The M3 type of events are found to be
similar to each other within quoted errors and lie between M1 and M2. This indicates
presence of localized independent fluctuations in both $N_{\gamma-{\rm like}}$ and 
$N_{\rm ch}$.

In summary, we have presented multiplicity distributions and multiplicity
fluctuations of photons and charged particles produced in
158$\cdot A$ GeV Pb+Pb collisions at the CERN-SPS. Multiplicity fluctuations
from data agree reasonably well with participant model calculations over
a wide range of centrality, from peripheral to central collisions.
This indicates the absence of any critical fluctuations indicative
of phase transition at SPS energies. Localized fluctuations in photons
and charged particles have been studied in order to search for formation
of domains of DCC. The results from correlation and DWT analysis are
compared to those of mixed events generated from data.
The results provide model independent evidence for non-statistical
fluctuations in the data for $\phi$ interval between 45 and 90 degrees
and one unit in pseudorapidity. This is seen to be due to individual
fluctuations in both photons and charged particles. No significant E-by-E correlated 
fluctuations are observed, contrary to expectations for a DCC effect.

\end{document}